\documentclass[a4paper,12pt]{article}
\pdfoutput=1
\usepackage{graphicx,subfigure,amsmath,amssymb}
\usepackage{color}
\usepackage{cite}
\usepackage{epstopdf}

\renewcommand{\baselinestretch}{1.5}
\newlength{\dinwidth}
\newlength{\dinmargin}
\begin{document}
\newcommand{\beq}{\begin{eqnarray}}
\newcommand{\eeq}{\end{eqnarray}}
%%%%%%%%%%%%%%%%%%%%%%%%%%%%%%%%%%%%%%%%%%%%%%%%%

\title{ Probing the R-parity violating supersymmetric effects in $B_c\to J/\psi\ell^- \bar{\nu}_{\ell},\eta_c\ell^- \bar{\nu}_{\ell}$  and
$\Lambda_b \to \Lambda_c \ell^- \bar{\nu}_{\ell}$ decays }
\author{\small Bin Wei$^{1}$%\thanks{E-mail:weibinwuli@163.com}
, ~Jie Zhu$^{1,2}$, ~Jin-Huan Sheng$^{1,3}$%\thanks{E-mail:jinhuanwuli@126.com}
, ~Ru-Min Wang$^{1}$\thanks{E-mail:ruminwang@sina.com}, ~Gong-Ru Lu$^{2}$
\\
  \renewcommand{\baselinestretch}{1.1}
{\scriptsize {$^1$ \it Institute of
    Theoretical Physics, Xinyang Normal University, Xinyang, Henan 464000, China
}}\\
{\scriptsize {$^2$ \it Institute of Particle and Nuclear Physics, Henan Normal University, Xinxiang, Henan 453007, China}}\\
{\scriptsize {$^3$ \it Institute of Particle Physics, Central China Normal University, Wuhan, Hubei 430079, China}}
}\renewcommand{\baselinestretch}{1.4}
\date{\today}
\maketitle

Motivated by recent $R_D$, $R_{D^*}$ and $R_{J/\psi}$ anomalies in $B\to D\ell^- \bar{\nu}_{\ell}$, $B\to D^*\ell^- \bar{\nu}_{\ell}$ and $B_c\to J/\psi\ell^- \bar{\nu}_{\ell}$ decays, respectively, we study possible R-parity violating supersymmetric effects in $B_c\to J/\psi\ell^- \bar{\nu}_{\ell},\eta_c\ell^- \bar{\nu}_{\ell}$  and
$\Lambda_b \to \Lambda_c \ell^- \bar{\nu}_{\ell}$ decays, which are also induced by $b\to c \ell^- \bar{\nu}_{\ell}$ at quark level. We find that (I) the constrained slepton exchange couplings $\lambda_{i33}\tilde{\lambda}'^*_{i23}$ involving in $b\to c \tau^- \bar{\nu}_{\tau}$ transition from relevant latest experimental data  still have quite large effects on  all (differential) branching ratios and the normalized forward-backward  asymmetries of  the exclusive  semileptonic $b\to c \tau^- \bar{\nu}_{\tau}$ decays as well as the ratios of the (differential) branching ratios;
%(II) the most of branching ratios of the exclusive semileptonic $b\to c \tau^- \bar{\nu}_{\tau}$ decays and ratios of the branching ratios  are very sensitive to $\lambda_{i33}\tilde{\lambda}'^*_{i23}$;
(II) after satisfying the data of $R_D$ and $R_{D^*}$,  the upper limit of $R_{J/\psi}$, $R_{\eta_c}$ and $R_{\Lambda_c}$ could be increased by 10\%, 112\% and 24\%, respectively,  from their upper limits of the Standard Model predictions by the $\lambda_{i33}\tilde{\lambda}'^*_{i23}$ couplings.
The results in this work could be used to
probe R-parity violating effects and will correlate with searches for direct supersymmetric signals at the
running LHCb and the forthcoming Belle-II.
\vspace{1.5cm}

\newpage

\section{Introduction}

Lepton flavor universality violation in the exclusive $b\to c\ell^-\bar{\nu}_\ell$ and $b\to s\ell^+\ell^-$ decays  has attracted a lot of attention in the
particle physics community and
has significantly constrained many  possible New Physics
(NP) effects. For ratios $R_{D^{(*)}}=\frac{\mathcal{B}(B\to D^{(*)}\tau^-\bar{\nu}_\tau)}{\mathcal{B}(B\to D^{(*)}\ell'^-\bar{\nu}_{\ell'})}$ with $\ell'=e$ or $\mu$,
the world average  of the BABAR \cite{Lees:2012xj,Lees:2013uzd}, Belle\cite{Sato:2016svk,
Huschle:2015rga,Hirose:2016wfn} and LHCb \cite{Aaij:2015yra,Aaij:2017uff} measurements  are \cite{HFAG}
\begin{eqnarray}
R_D^{Exp.} &=&0.407\pm 0.039\pm 0.024,\nonumber\\
R_{D^{\ast}}^{Exp.} &=& 0.304\pm 0.013\pm 0.007,
\end{eqnarray}
which are exceed the Standard Model (SM) predictions ($R_D^{SM}=0.297\pm 0.017$ \cite{Na:2015kha}, $R_{D^{\ast}}^{SM}=0.252\pm 0.003$ \cite{Fajfer:2012vx}) by $1.9\sigma$ and $3.3\sigma$, respectively.   Very recently, LHCb reported a new measurement regarding $b\to c \ell\nu$ in $B_c$ decays \cite{Aaij:2017tyk}
\begin{eqnarray}
R_{J/\psi}=\frac{\mathcal{B}(B^+_c\to  J/\psi\tau^+\nu_\tau)}{\mathcal{B}(B^+_c\to  J/\psi\mu^+\nu_\mu)}=0.71\pm0.17\pm0.18,
\end{eqnarray}
which is about $2\sigma$ higher than its SM prediction \cite{Wen-Fei:2013uea,Dutta:2017xmj}.

In the ratios $R_D$, $R_{D^*}$ and  $R_{J/\psi}$, the theoretical uncertainties, such as the relevant CKM matrix elements and  form factors, are largely  canceled, so any deviation from the SM prediction would clearly indicate the presence of NP. A lot of works about $R_D$ and $R_{D^*}$  have been done to explain these anomalies in different NP models, for examples,
model-independent approaches \cite{Choudhury:2017qyt,Dutta:2017wpq,Ivanov:2017mrj,Bardhan:2016uhr,Faroughy:2016osc,Feruglio:2016gvd,Bhattacharya:2015ida,Sakaki:2014sea,Sakaki:2013bfa}, charged Higgs \cite{Iguro:2017ysu,Celis:2016azn,Crivellin:2015hha,Kim:2015zla,Crivellin:2013wna,Celis:2012dk}, lepton flavor violation \cite{Barbieri:2015yvd,Freytsis:2015qca,Alonso:2015sja,Ko:2012sv},  leptoquark \cite{Sahoo:2016pet,Li:2016vvp,Hati:2015awg,Calibbi:2015kma}, etc.  Otherwise, NP effects in  $B_c\to J/\psi\ell^- \bar{\nu}_{\ell}$ and  $\Lambda_b \to \Lambda_c \ell^- \bar{\nu}_{\ell}$ decays have also been studied, for instance, in Refs. \cite{Watanabe:2017mip,Chauhan:2017uil,Dutta:2017xmj} and  Refs. \cite{Dutta:2015ueb,Li:2016pdv,Celis:2016azn,Faustov:2016pal,
Detmold:2015aaa,Gutsche:2015mxa,Shivashankara:2015cta}, respectively.

In the supersymmetry without R-parity, the slepton exchange couplings could give  large contributions to $R_D$ and $R_{D^*}$, which we have studied in Ref. \cite{Zhu:2016xdg}.
 Since $B_c\to J/\psi\ell^- \bar{\nu}_{\ell},\eta_c\ell^- \bar{\nu}_{\ell}$  and
$\Lambda_b \to \Lambda_c \ell^- \bar{\nu}_{\ell}$ decays are also induced  at quark level by $b\to c \ell^- \bar{\nu}_{\ell}$, they involve the same set of R-parity violating (RPV) coupling constants as $B\to D\ell^- \bar{\nu}_{\ell}$ and $B\to D^*\ell^- \bar{\nu}_{\ell}$ decays. In this work, using the latest experimental data of all relevant exclusive $b\to c \ell^- \bar{\nu}_{\ell}$ decays, we will analyze the constrained RPV contributions  to the
branching ratios and their ratios, differential branching ratios as well as normalized forward-backward (FB) asymmetries of the charged
leptons in $B_c\to J/\psi\ell^- \bar{\nu}_{\ell},\eta_c\ell^- \bar{\nu}_{\ell}$  and
$\Lambda_b \to \Lambda_c \ell^- \bar{\nu}_{\ell}$ decays.

The paper is organized as follows. In section 2, we briefly review the theoretical expressions
of the exclusive $b \to c \ell^- \bar{\nu}_\ell$
decays. In section 3, using the constrained parameter spaces from relevant experimental measurements, we make a detailed classification research on the RPV effects on the quantities which have not been measured or not been well measured yet. Our conclusions are given in section 4.

\section{Theoretical Framework}

The general effective Hamiltonian for $b \to c \ell_m^- \bar{\nu}_{\ell_n}$ transitions can be written as \cite{Bhattacharya:2011qm,Cirigliano:2009wk}
\begin{eqnarray}
\mathcal{H}_{eff}(b \to c \ell_m^- \bar{\nu}_{\ell_n})&=& \frac{G_FV_{cb}}{\sqrt{2}}\Bigg\{~~\Big[G_V\bar{c}\gamma_{\mu}b-G_A\bar{c}\gamma_{\mu}\gamma_{5}b\Big]\bar{\ell}_m^-\gamma^{\mu}(1-\gamma_5)\bar{\nu}_{\ell_n}~~~~~~~~~~~~~~~~~~~~\nonumber\\
&& ~~~~~~~~~+\Big[G_S\bar{c}b-G_P\bar{c}\gamma_{5}b\Big]\bar{\ell}_m^-(1-\gamma_5)\bar{\nu}_{\ell_n}\nonumber\\
&& ~~~~~~~~~+\Big[\widetilde G_V \bar{c}\gamma_{\mu}b-\widetilde G_A\bar{c}\gamma_{\mu}\gamma_5 b\Big]\bar{\ell}_m^-\gamma^{\mu}(1+\gamma_5)\bar{\nu}_{\ell_n}\nonumber\\
&& ~~~~~~~~~+\Big[\widetilde G_S\bar{c}b-\widetilde G_P\bar{c}\gamma_5 b\Big]\bar{\ell}_m^-(1+\gamma_5)\bar{\nu}_{\ell_n}   \Bigg\}+h.c.\label{Eq:Heff}
\end{eqnarray}
In the SM, $G_V=G_A=1$ and all others are zero. If considering both the SM and the RPV contributions,  we have \cite{Kim:2007uq}
\begin{eqnarray}
G_V=G_A&=&1-\frac{\sqrt{2}}{G_FV_{cb}}\sum_i\frac{\lambda'_{n3i}\tilde{\lambda}'_{m2i}}{8m^2_{\tilde{d}_{iR}}},\\
G_S=G_P&=&\frac{\sqrt{2}}{G_FV_{cb}}\sum_i\frac{\lambda_{inm}\tilde{\lambda}'_{i23}}{4m^2_{\tilde{\ell}_{iL}}},
\end{eqnarray}
and all others are zero.

From general  effective Hamiltonian given in Eq. (\ref{Eq:Heff}), we obtain the
differential branching ratios.
The detail expressions  can be found in   \ref{A:B} and \ref{A:Lambdab} of Appendix. We only give the final ones in this subsection.
For $B_c\to \eta_c  \ell^- \bar{\nu}_\ell$ and $B\to D  \ell^- \bar{\nu}_\ell$ decays,
\begin{eqnarray}
\frac{d\mathcal{B}(B_q\to P \ell^- \bar{\nu}_\ell)}{dq^2} &=& \frac{G_F^2|V_{cb}|^2\tau_{B_q}|\vec{p}_P|q^2}{96\pi^3m_{B_q}^2}\left(1-\frac{m_\ell^2}{q^2}\right)^2\Biggl\{H_0^2(\big|G_V\big|^2+ \big|\widetilde G_V\big|^2)\Big(1+\frac{m_\ell^2}{2q^2}\Big)\nonumber\\
&& +\frac{3m_\ell^2}{2q^2}\Big[\Big|H_tG_V+\frac{\sqrt{q^2}}{m_\ell}H_SG_S\Big|^2+\Big|H_t\widetilde G_V+\frac{\sqrt{q^2}}{m_\ell}H_S\widetilde G_S\Big|^2\Big]\Biggl\},
\end{eqnarray}
with
\begin{eqnarray}
H_0&=& \frac{2m_{B_q}|\vec{p}_P|}{\sqrt{q^2}}F_+(q^2),\nonumber\\
H_t&=& \frac{m_{B_q}^2-m_P^2}{\sqrt{q^2}}F_0(q^2),\nonumber\\
H_S&=& \frac{m_{B_q}^2-m_P^2}{m_b-m_{c}}F_0(q^2).
\end{eqnarray}

For $B_c\to J/\psi  \ell^- \bar{\nu}_\ell$ and $B\to D^*  \ell^- \bar{\nu}_\ell$ decays,
\begin{eqnarray}
\frac{d\mathcal{B}(B_q\to V \ell^- \bar{\nu}_\ell)}{dq^2}&=&\frac{G_F^2|V_{cb}|^2\tau_{B_q}|\vec{p}_V|q^2}{96\pi^3m_{B_q}^2}\left(1-\frac{m_\ell^2}{q^2}\right)^2\Biggl\{\big|\mathcal{A}_{AV}\big|^2
+\frac{m_\ell^2}{2q^2}\Big(\big|\mathcal{A}_{AV}\big|^2+3\big|\mathcal{A}_{tP}\big|^2\Big)\nonumber\\
&&+\big|\widetilde{\mathcal{A}}_{AV}\big|^2+\frac{m_\ell^2}{2q^2}\Big(\big|\widetilde{\mathcal{A}}_{AV}\big|^2+3\big|\widetilde{\mathcal{A}}_{tP}\big|^2\Big)\Biggl\},
\end{eqnarray}
where
\begin{eqnarray}
&&\Big|\mathcal{A}_{AV}\Big|^2=\mathcal{A}^2_0\Big|G_A\Big|^2+\mathcal{A}^2_{\|}\Big|G_A\Big|^2+\mathcal{A}^2_{\perp}\Big|G_V\Big|^2,\nonumber\\
&&\Big|\widetilde{\mathcal{A}}_{AV}\Big|^2=\mathcal{A}^2_0\Big|\widetilde{G}_A\Big|^2+\mathcal{A}^2_{\|}\Big|\widetilde{G}_A\Big|^2+\mathcal{A}^2_{\perp}\Big|\widetilde{G}_V\Big|^2,\nonumber\\
&&\Big|\mathcal{A}_{tP}\Big|^2=\Big|\mathcal{A}_0G_A+\frac{\sqrt{q^2}}{m_\ell}\mathcal{A}_PG_P\Big|^2,\nonumber\\
&&\Big|\widetilde{\mathcal{A}}_{tP}\Big|^2=\Big|\mathcal{A}_0\widetilde{G}_A+\frac{\sqrt{q^2}}{m_\ell}\mathcal{A}_P\widetilde{G}_P\Big|^2,
\end{eqnarray}
with
\begin{eqnarray}\label{B2Vtransition2}
\mathcal{A}_0&=& \frac{1}{2m_V\sqrt{q^2}}\left[(m_{B_q}^2-m_V^2-q^2)(m_{B_q}+m_V)A_1(q^2)-\frac{4m_{B_q}^2|\vec{p}_V|^2}{m_{B_q}+m_V}A_2(q^2)\right],\nonumber\\
\mathcal{A}_{\|}&=& \sqrt{2}(m_{B_q}+m_V)A_1(q^2),\nonumber\\
\mathcal{A}_{\perp}&=& -\frac{4m_{B_q}V(q^2)|\vec{p}_V|}{\sqrt{2}(m_{B_q}+m_V)},\nonumber\\
\mathcal{A}_t&=& \sqrt{2}m_{B_q}|\vec{p}_V|A_0(q^2),\nonumber\\
\mathcal{A}_P&=& -\frac{2m_{B_q}|\vec{p}_V|A_0(q^2)}{m_b+m_c}.
\end{eqnarray}

For baryonic $\Lambda_b \to \Lambda_c \ell^- \bar{\nu}_\ell$ decays,
\begin{equation}
\frac{d\mathcal{B}(\Lambda_b \to \Lambda_c \ell^- \bar{\nu}_\ell)}{dq^2}=\frac{G_F^2|V_{cb}|^2\tau_{\Lambda_b}|\vec{p}_{\Lambda_c}|q^2}{192\pi^3m^2_{\Lambda_b}}\left(1-\frac{m_\ell^2}{q^2}\right)^2\left[B_1+\frac{m_\ell^2}{2q^2}B_2+\frac{3}{2}B_3+\frac{3m_\ell}{\sqrt{q^2}}B_4\right],
\end{equation}
where
\begin{eqnarray}\label{Lb2Lc12}
B_1 &=& \Big|H_{\frac{1}{2} 0}\Big|^2+\Big|H_{-\frac{1}{2} 0}\Big|^2+\Big|H^2_{\frac{1}{2} 1}\Big|^2+\Big|H_{-\frac{1}{2} -1}\Big|^2,\nonumber\\
B_2 &=& \Big|H_{\frac{1}{2} 0}\Big|^2+\Big|H_{-\frac{1}{2} 0}\Big|^2+\Big|H_{\frac{1}{2} 1}\Big|^2+\Big|H_{-\frac{1}{2} -1}\Big|^2+3\Big(\Big|H_{\frac{1}{2} t}\Big|^2+\Big|H_{-\frac{1}{2} t}\Big|^2\Big),\nonumber\\
B_3 &=& \Big|H^{SP}_{\frac{1}{2} 0}\Big|^2+\Big|H^{SP}_{-\frac{1}{2} 0}\Big|^2,\nonumber\\
B_4 &=& Re\Big[H_{\frac{1}{2} t}H^{SP*}_{\frac{1}{2} 0}+H_{-\frac{1}{2} t}H^{SP*}_{-\frac{1}{2} 0}\Big],
\end{eqnarray}
with
\begin{eqnarray}
H_{\lambda_2\lambda_1}&\equiv&H^V_{\lambda_2\lambda_1}-H^A_{\lambda_2\lambda_1},\nonumber\\
H^V_{\frac{1}{2}0}&=&G_V\frac{\sqrt{Q_-}}{\sqrt{q^2}}[(m_{\Lambda_b}+m_{\Lambda_c})f_1{q^2}-q^2f_2(q^2)],\nonumber\\
H^A_{\frac{1}{2}0}&=&G_A\frac{\sqrt{Q_+}}{\sqrt{q^2}}[(m_{\Lambda_b}-m_{\Lambda_c})g_1{q^2}+q^2g_2(q^2)],\nonumber\\
H^V_{\frac{1}{2}1}&=&G_V\sqrt{2Q_-}[-f_1{q^2}+(m_{\Lambda_b}+m_{\Lambda_c})f_2(q^2)],\nonumber\\
H^A_{\frac{1}{2}1}&=&G_A\sqrt{2Q_+}[-g_1{q^2}-(m_{\Lambda_b}-m_{\Lambda_c})g_2(q^2)],\nonumber\\
H^V_{\frac{1}{2}t}&=&G_V\frac{\sqrt{Q_+}}{\sqrt{q^2}}[(m_{\Lambda_b}-m_{\Lambda_c})f_1{q^2}+q^2f_3(q^2)],\nonumber\\
H^A_{\frac{1}{2}t}&=&G_A\frac{\sqrt{Q_-}}{\sqrt{q^2}}[(m_{\Lambda_b}+m_{\Lambda_c})g_1{q^2}-q^2g_3(q^2)],\\
H^{SP}_{\frac{1}{2}0} &\equiv& H^S_{\frac{1}{2}0}-H^P_{\frac{1}{2}0},\nonumber\\
H^S_{\frac{1}{2} 0} &=& G_S\frac{\sqrt{Q_+}}{m_b-m_{c}}[(m_{\Lambda_b}-m_{\Lambda_c})f_1(q^2)+q^2f_3(q^2)],\nonumber\\
H^P_{\frac{1}{2} 0} &=& G_P\frac{\sqrt{Q_-}}{m_b+m_{c}}[(m_{\Lambda_b}+m_{\Lambda_c})g_1(q^2)-q^2g_3(q^2)],
\end{eqnarray}
where $Q_\pm =(m_{\Lambda_b} \pm m_{\Lambda_c})^2-q^2$.  Either from parity or from explicit calculation, we have the relations $H^V_{-\lambda_2 -\lambda_1}=H^V_{\lambda_2 \lambda_1}$, $H^A_{-\lambda_2 -\lambda_1}=-H^A_{\lambda_2 \lambda_1}$, $H^S_{\lambda_2\lambda_1}=H^S_{-\lambda_2-\lambda_1}$ and $H^P_{\lambda_2\lambda_1}=-H^P_{-\lambda_2-\lambda_1}$.

In order to further study the RPV effects, we need calculate other two important physical
quantities in  $ M_1\rightarrow M_2{\ell^-}\bar{\nu}_{\ell}$ decays  to reduce the error.
The ratio of  differential branching ratio may be written as
\begin{eqnarray}
\frac{dR_{M_2}}{dq^2}&=&\frac{d\Gamma(M_1\rightarrow M_2{{\tau}^-}\bar{\nu}_\tau)/ds}{d\Gamma(M_1\rightarrow M_2{\ell'^-}\bar{\nu}_{\ell'})/ds}.
\end{eqnarray}
Noted that $R_{M_2}$ is obtained by  separately integrating  the numerators  and denominators of above $dR_{M_2}/dq^2$.
The normalized forward-backward asymmetry is defined as
\begin{equation}
A_{FB}^{M_1\rightarrow M_2{\ell^-}\bar{\nu}_{\ell}}(q^2)=\frac{\int^0_{-1}dcos{\theta_\ell}\Big[\frac{d^2\Gamma(M_1\rightarrow M_2{\ell^-}\bar{\nu}_{\ell})}{dq^2dcos{\theta_\ell}}\Big]-\int^1_{0}dcos{\theta_\ell}\Big[\frac{d^2\Gamma(M_1\rightarrow M_2{\ell^-}\bar{\nu}_{\ell})}{dq^2dcos{\theta_\ell}}\Big]}
{  \frac{d\Gamma(M_1\rightarrow M_2{\ell^-}\bar{\nu}_{\ell})}{dq^2}}.
\end{equation}

%\clearpage
\section{Numerical Results and Discussions}
The main theoretical input parameters are the transition form factors, the CKM matrix element $V_{cb}$, the masses, the mean lives,  etc.
Relevant transition form factors are taken from Refs. \cite{Caprini:1997mu,Amhis:2016xyh,Wen-Fei:2013uea,Detmold:2015aaa},  the CKM matrix element is taken from the UTfit Collaboration \cite{UTFit}, and others are gotten  from PDG \cite{Olive:2016xmw}. The 95\% confidence level (CL) theoretical uncertainties of the input parameters are considered in our results.

In our calculation, we consider only one NP coupling at one time and keep its interference with the SM amplitude to study the RPV  effects.
Due to the strong helicity suppression, the squark exchange couplings have no very obvious effects on the differential branching ratios and the normalized FB asymmetries of the semileptonic  exclusive $b \to c \ell^- \bar{\nu}_{\ell}$  decays. So we will only focus on the slepton exchange couplings in our following discussions.
We assume the masses of the corresponding slepton are 500 GeV, for other values of the slepton masses, the bounds on the couplings in this
paper can be easily obtained by scaling them by factor of $\tilde{f}^2\equiv\left(\frac{m_{\tilde{\ell}}}{500\rm{GeV}}\right)^2$.

 A part of latest relevant experimental ranges at 95\% CL are listed in the second column of Tab. \ref{tab:sleptonBr&R}. The following experimental constraints at 95\% CL of $B_u$ decays will be also considered in our analysis.
\begin{eqnarray}
\mathcal{B}(B_u  \to D^{\ast 0} \ell'^- \bar{\nu_{\ell'}})&=&(5.69\pm0.19)\times10^{-2},\nonumber\\
\mathcal{B}(B_u  \to D^{\ast 0} \tau^- \bar{\nu_{\tau}})&=&(1.88\pm0.20)\times10^{-2},\nonumber\\
\mathcal{B}(B_u  \to D \ell'^- \bar{\nu_{\ell'}})&=&(2.27\pm0.11)\times10^{-2},\nonumber\\
\mathcal{B}(B_u  \to D \tau^- \bar{\nu_{\tau}})&=&(7.7\pm2.5)\times10^{-3}.
\end{eqnarray}
Noted that the slepton exchange contributions to $B^-_u\to D^{(*)0}_u\ell^-\nu_{\ell}$ and $B^0_d\to D^{(*)+}_d\ell^-\nu_{\ell}$  are very similar to each other, since  the $SU(2)$ flavor symmetry implies $M(B^-_u\to D^{(*)0}_u\ell^-\nu_{\ell})\approx M(B^0_d\to D^{(*)+}_d\ell^-\nu_{\ell})$.
So we would take $B^0_d\to D^{(*)+}_d\ell^-\nu_{\ell}$  decays as examples in the following.
Since the experimental measurements  of $\mathcal{B}(B \to D^* \tau^- \bar{\nu}_{\tau})$ and $\mathcal{R}(D^*)$ obviously deviate from their SM predictions,  we do not impose some obvious deviated measurements and just leave them as predictions
of the restricted parameter spaces of the RPV couplings,
and then compare them with the experimental results. Two schemes of 95\% CL experimental bounds will be used in this work.
\begin{itemize}
\item[$S_1:$] All relevant experimental bounds except for $\mathcal{B}(B \to D^* \tau^- \bar{\nu}_{\tau})$ and  $\mathcal{R}(D^*)$ at $95\%$ CL.
\item[$S_2:$] All relevant experimental bounds except for $\mathcal{B}(B
 \to D^* \tau^- \bar{\nu}_{\tau})$ at $95\%$ CL.
\end{itemize}

\begin{figure}[b]
\begin{center}
\includegraphics[scale=0.6]{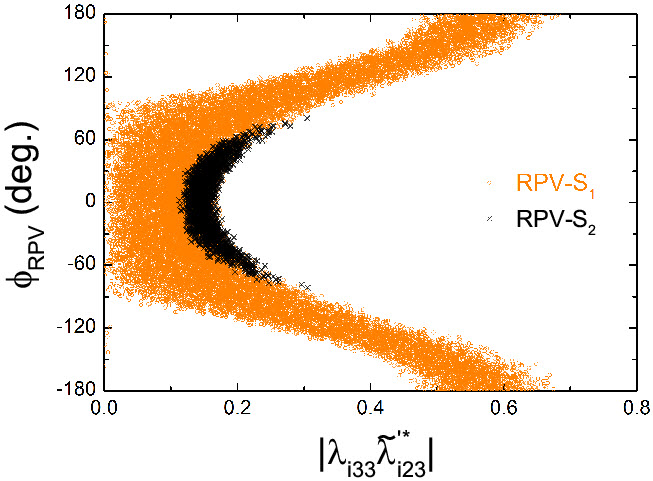}
\end{center}
\caption{The allowed spaces due to slepton exchange couplings from the latest 95\% CL experimental bounds of the exclusive $b\to c \ell^-\bar{\nu}_\ell$ decays. }
\label{fig:sleptonspace}
\end{figure}

Slepton exchange couplings $\lambda_{i11}\tilde{\lambda}'^*_{i23}$, $\lambda_{i22}\tilde{\lambda}'^*_{i23}$ and $\lambda_{i33}\tilde{\lambda}'^*_{i23}$  contribute to the exclusive $b\to c e^-\bar{\nu}_e$, $b\to c \mu^-\bar{\nu}_\mu$ and  $b\to c \tau^-\bar{\nu}_\tau$ decays, respectively.
For $\lambda_{i11}\tilde{\lambda}'^*_{i23}$  and  $\lambda_{i22}\tilde{\lambda}'^*_{i23}$ , which contribute to both $b\to c \ell'^- \bar{\nu}_{\ell'}$ and $b\to s \ell'^+ \ell'^-$ transitions, the stronger constraints come from the exclusive  $b\to s \ell'^+ \ell'^-$ decays~( $|\lambda_{i11}\tilde{\lambda}'^*_{i23}|<5.75\times10^{-4}$,  $|\lambda_{i22}\tilde{\lambda}'^*_{i23}|<1.63\times10^{-5}$ )\cite{Xu:2006vk,Wang:2011aa}, which  will be used in our numerical results.
Fig.\ref{fig:sleptonspace} shows the allowed coupling spaces of  $\lambda_{i33}\tilde{\lambda}'^*_{i23}$ from the latest 95\% CL experimental measurements of the exclusive $b\to c \tau^-\bar{\nu}_\tau$ in the cases of $S_1$ and $S_2$.
Fig.\ref{fig:sleptonspace}  shows us that both  moduli and RPV weak phases of $\lambda_{i33}\tilde{\lambda}'^*_{i23}$ are constrained in the cases of both $S_1$ and $S_2$. In the  $S_1$, we obtain $|\lambda_{i33}\tilde{\lambda}'^*_{i23}|\leq 0.67$, and the slight difference between this constrained space within $S_1$ case and one in Fig. 3 of Ref. \cite{Zhu:2016xdg}  comes from the updated input parameters and experimental measurements. If considering the $R_{D^*}$ bound, $i.e.$, in the cases of $S_2$, very strong bounds on $\lambda_{i33}\tilde{\lambda}'^*_{i23}$ are obtained, $0.11\leq|\lambda_{i33}\tilde{\lambda}'^*_{i23}|\leq 0.30$ and $|\phi_{RPV}|\leq 82^\circ$.

\begin{figure}[t]
\begin{center}
\includegraphics[scale=1.12]{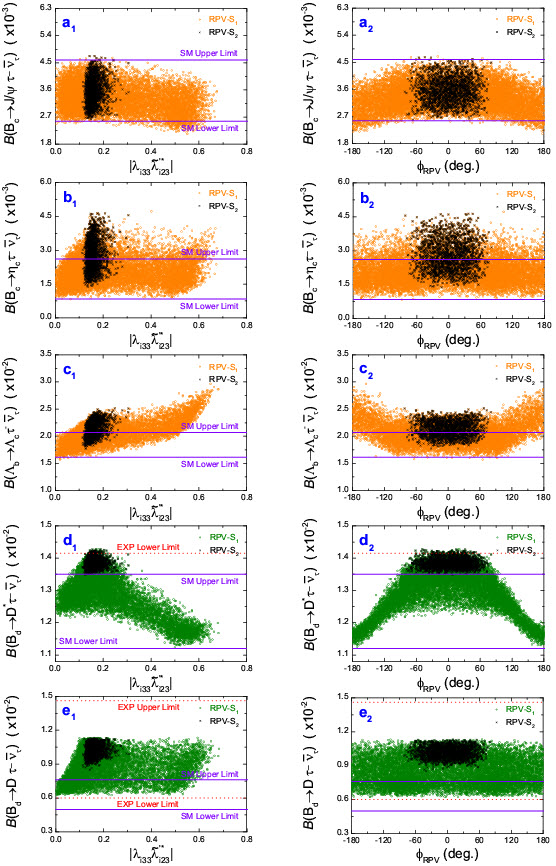}
\end{center}\vspace{-0.8cm}
 \caption{The constrained slepton exchange coupling effects on the branching ratios of the exclusive semileptonic $b\to c \tau^-\bar{\nu}_\tau$ decays. }
\label{fig:BR}
\end{figure}

\begin{figure}[t]
\begin{center}
\includegraphics[scale=1.15]{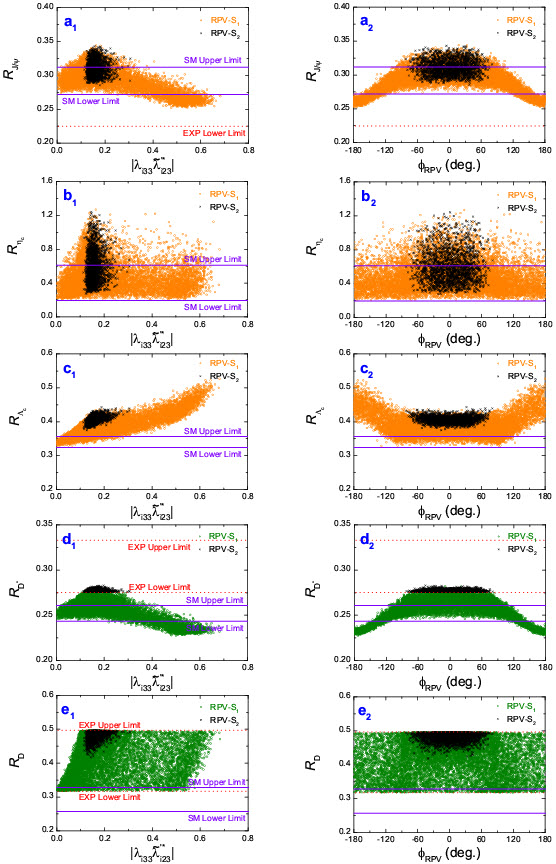}
\end{center}\vspace{-0.8cm}
 \caption{The constrained slepton exchange coupling effects on $R_{J/\psi}$, $R_{\eta_c}$, $R_{\Lambda_c}$, $R_{D^*}$ and $R_{D}$. }
\label{fig:R}
\end{figure}

Now we discuss the constrained slepton exchange effects in $B_c\to J/\psi\ell^- \bar{\nu}_{\ell},\eta_c\ell^- \bar{\nu}_{\ell}$  and
$\Lambda_b \to \Lambda_c \ell^- \bar{\nu}_{\ell}$ decays.
Our numerical predictions for the  branching ratios and their ratios are summarized in last columns of  Tab. \ref{tab:sleptonBr&R}. We also show their sensitivities to the moduli and weak phases of the slepton exchange couplings $\lambda_{i33}\tilde{\lambda}'^*_{i23}$ in Figs. \ref{fig:BR}-\ref{fig:R}.
\begin{table}[b]
\caption{Experimental ranges and our numerical predictions for the branching ratios (in units of $10^{-2}$) and their ratios.}
\begin{center}
\begin{tabular}
{lcccc}\hline\hline
Observable & $Exp.$ ranges & SM predictions & RPV-$S_1$ & RPV-$S_2$ \\\hline
$\mathcal{B}(B_c \to J/\psi \ell'^- \bar{\nu}_{\ell'})$ & $\cdot\cdot\cdot$ & $[~0.86~,~1.67~]$ & $[~0.77~,~1.64~]$ & $[~0.77~,~1.61~]$\\
$\mathcal{B}(B_c \to J/\psi \tau^- \bar{\nu}_{\tau}) $ & $\cdot\cdot\cdot$ & $[~0.26~,~0.46~]$ & $[~0.22~,~0.47~]$ &  $[~0.26~,~0.47~]$\\\hline
$\mathcal{B}(B_c \to \eta_c \ell'^- \bar{\nu}_{\ell'})$ & $\cdot\cdot\cdot$ & $[~0.27~,~0.88~]$ & $[~0.25~,~0.87~]$ & $[~0.25~,~0.83~]$\\
$\mathcal{B}(B_c \to \eta_c \tau^- \bar{\nu}_{\tau})$ & $\cdot\cdot\cdot$ & $[ 0.084,~0.263]$ & $[0.087,~0.478]$ & $[0.138,~0.464]$ \\\hline
$\mathcal{B}(\Lambda_b \to \Lambda_c \ell'^- \bar{\nu}_{\ell'})$ & $\cdot\cdot\cdot$ & $[~4.75~,~6.33~]$ & $[~4.38~,~6.22~]$ & $[~4.38~,~6.22~]$\\
$\mathcal{B}(\Lambda_b \to \Lambda_c {\tau}^- \bar{\nu}_{\tau})$ & $\cdot\cdot\cdot$ & $[~1.60~,~2.08~]$ & $[~1.57~,~2.97~]$ &  $[~1.77~,~2.52~]$\\\hline
$\mathcal{B}(B_d  \to D_d^{\ast } \ell'^- \bar{\nu}_{\ell'})$ & $ [~4.71~,~\textcolor[rgb]{0,0,1.00}{5.15~}]$ & $[~4.45~,~5.32~]$ & $[~4.88~,~\textcolor[rgb]{0,0,1.00}{5.15~}]$ & $[~4.88~,~\textcolor[rgb]{0,0,1.00}{5.15~}]$\\
$\mathcal{B}(B_d \to D_d^{\ast } {\tau}^- \bar{\nu}_{\tau})$ & $ [~\textcolor[rgb]{1.00,0,0}{1.42~},~1.92
~]$ & $[~1.12~,~1.35~]$ & $[~1.12~,~\textcolor[rgb]{1.00,0,0}{1.43~}]$ & $[~1.35~,~\textcolor[rgb]{1.00,0,0}{1.43~}]$ \\\hline
$\mathcal{B}(B_d \to D_d \ell'^- \bar{\nu}_{\ell'})$ & $ [~\textcolor[rgb]{0,0,1.00}{1.95~},~2.43~]$ & $ [~1.82~,~2.46~]$ & $[~\textcolor[rgb]{0,0,1.00}{1.95~},~2.32~]$ & $[~\textcolor[rgb]{0,0,1.00}{1.95~},~2.32~]$ \\
$\mathcal{B}(B_d \to D_d {\tau}^- \bar{\nu}_{\tau})$ & $[~0.60~,~1.46~]$ & $ [~0.50~,~0.76~]$ & $[~0.62~,~1.14~]$ & $[~0.85~,~1.14~]$ \\\hline
$R_{J/\psi}$ & $ [0.225,~1.195]$ & $[0.271,~0.314] $ & $[0.252,~0.346]$ & $[0.285,~0.346]$ \\
$R_{\eta_c}$ & $\cdot\cdot\cdot$ & $[0.192,~0.613]$ & $[0.204,~1.300]$ & $[0.274,~1.300]$ \\
$R_{\Lambda_c}$ & $\cdot\cdot\cdot$ & $ [0.322,~0.356]$ & $[0.328,~0.527]$ & $[0.373,~0.443]$ \\
$R_D$ & $[\textcolor[rgb]{0,0,1.00}{0.317},~0.497]$ & $[0.255,~0.328]$ & $[\textcolor[rgb]{0,0,1.00}{0.317},~\textcolor[rgb]{0,0,1.00}{0.497}]$& $[0.414,~\textcolor[rgb]{0,0,1.00}{0.497}]$ \\
$R_{D^{\ast}}$ & $[\textcolor[rgb]{1.00,0,0}{0.275},~0.333]$ & $ [0.242,~0.262]$ & $[0.226,~\textcolor[rgb]{1.00,0,0}{0.283}]$& $[0.275,~\textcolor[rgb]{1.00,0,0}{0.283}]$\\\hline\hline
\end{tabular}
\end{center}
\label{tab:sleptonBr&R}
\end{table}
For convenient analysis and comparison, we also give all SM predictions of $B_c\to J/\psi\ell^- \bar{\nu}_{\ell},\eta_c\ell^- \bar{\nu}_{\ell}$  and
$\Lambda_b \to \Lambda_c \ell^- \bar{\nu}_{\ell}$ decays   as well as  the similar updated  predictions of $B\to D^{(*)}\ell^-\bar{\nu}_\ell$ decays, which have been given  in Ref. \cite{Zhu:2016xdg}.
We have the following remarks for the branching ratios and their ratios:
\begin{itemize}

\item {\bf Experimental constaints}: In the case of $S_1$, the 95\% CL experimental upper limits of $\mathcal{B}(B_d\to
   D^{*}\ell'^-\nu_{\ell'})$ and $R_D$ as well as lower limits  of  $\mathcal{B}(B_d\to
   D\ell'^-\nu_{\ell'})$ and $R_D$  give effective constraints on the RPV couplings.    In the case of $S_2$, the lower limit of $R_{D^*}$ can give further effective constraint.

\item {\bf Branching ratios with $\ell=\ell'$}: The constrained slepton couplings have but not large effects on $\mathcal{B}(B_c \to J/\psi
   \ell'^- \bar{\nu}_{\ell'},\eta_c \ell'^- \bar{\nu}_{\ell'})$ and $\mathcal{B}(\Lambda_b \to
   \Lambda_c \ell'^- \bar{\nu}_{\ell'})$.  And these branching ratios are not very sensitive to relevant slepton
   exchange couplings, so we will not display their sensitivities to RPV  couplings as similar as Fig.
   \ref{fig:BR}.

\item {\bf Branching ratios with $\ell=\tau$}: As displayed in  Fig. \ref{fig:BR},  the constrained slepton couplings have  very obvious effects on all five branching ratios with $\ell=\tau$, and they are very sensitive to both moduli and weak phases of  $\lambda_{i33}\tilde{\lambda}'^*_{i23}$. If also considering the experimental bounds of $R_{D^*}$, the lower limits of $\mathcal{B}(B_c\to J/\psi\tau^- \bar{\nu}_{\tau},\eta_c\tau^- \bar{\nu}_{\tau})$  and $\mathcal{B}(B \to D^{*} \tau^- \bar{\nu}_\tau,D \tau^- \bar{\nu}_\tau)$   as well as both upper and lower limits of $\mathcal{B}(\Lambda_b \to \Lambda_c {\tau}^- \bar{\nu}_{\tau})$ are further constrained. As for $\mathcal{B}(B_d \to D_d^{\ast } {\tau}^- \bar{\nu}_{\tau})$, which experimental constraints are not used, within both $S_1$ and $S_2$ cases,  only very narrow ranges, $[1.42,1.43]\times10^{-2}$, satisfy  its present measurement.

\item {\bf Ratios of the branching ratios}: As shown in  Fig. \ref{fig:R}, the ratios of the branching ratios are also very sensitive to  $\lambda_{i33}\tilde{\lambda}'^*_{i23}$ couplings, the further experimental constraints of  $R_{D^*}$ give obvious lower limits of $R_{J/\psi}$, $R_{\Lambda_c}$ and $R_D$.  Present experimental measurement of $R_{J/\psi}$ with the large uncertainty  could not give any further constraint on the slepton exchange couplings. The upper limit of $R_{J/\psi}$, $R_{\eta_c}$ and $R_{\Lambda_c}$ could be increased by 10\%, 112\% and 24\%, respectively,  from their upper limits of the SM predictions. The upper limit of RPV prediction of $R_{J/\psi}$ is about half of the central value of the experimental measurement, but is within 1.5$\sigma$.

\end{itemize}

\begin{figure}[bh]
\begin{center}
\includegraphics[scale=1.1]{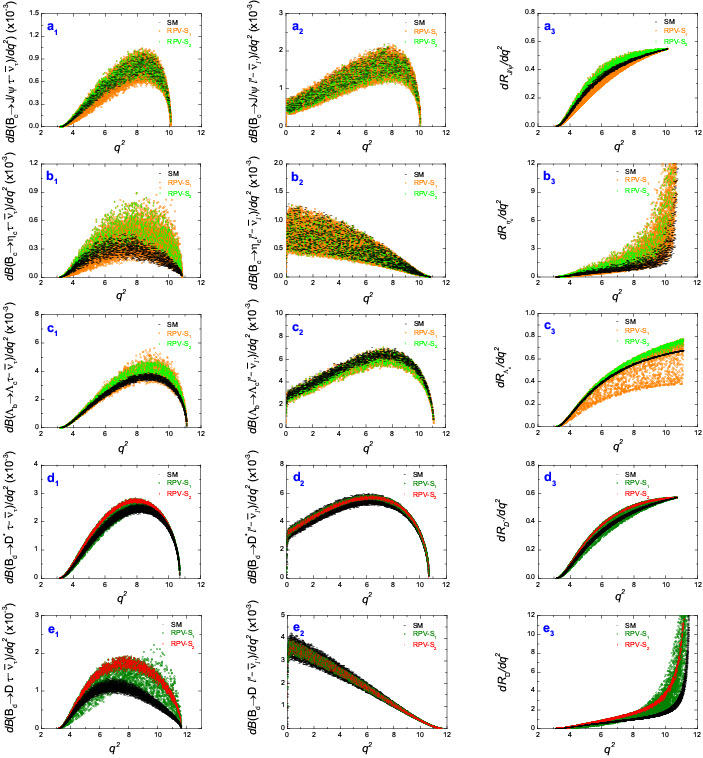}
\end{center}\vspace{-0.8cm}
 \caption{The constrained slepton exchange coupling effects on the differential branching ratios and their ratios of the exclusive semileptonic $b\to c \ell^-\bar{\nu}_\ell$ decays.}
\label{fig:dBrR}
\end{figure}

Now we discuss the constrained slepton exchange coupling effects on the differential branching ratios and their ratios of the exclusive semileptonic $b\to c \ell^-\bar{\nu}_\ell$ decays, which are shown in Fig. \ref{fig:dBrR}.
From  the first and last columns of Fig.\ref{fig:dBrR}, one can see that the constrained slepton exchange couplings have very large effects on all five  differential branching ratios with $\ell=\tau$ and all five ratios of  the differential branching ratios.  In $S_2$ case,   the 95\% CL experimental measurement of  $R_{D^*}$  given obviously  further constraints on the  lower limits of $d\mathcal{B}(B_c\to J/\psi \tau\nu_\tau)/dq^2$, $d\mathcal{B}(B_c\to \eta_c \tau\nu_\tau)/dq^2$,  $d\mathcal{B}(B \to D^* \tau\nu_\tau)/dq^2$, $dR_{J/\psi}/dq^2$,  $dR_{\eta_c}/dq^2$, $dR_{\Lambda_c}/dq^2$, $dR_{D^*}/dq^2$   as well as  both upper and lower limits of  $d\mathcal{B}(\Lambda_b\to \Lambda_c \tau\nu_\tau)/dq^2$, $d\mathcal{B}(B \to D \tau\nu_\tau)/dq^2$ and  $dR_{D}/dq^2$.
From  the second column of Fig.\ref{fig:dBrR},   one can see that the constrained slepton exchange couplings still have some effects on $d\mathcal{B}(B_c\to J/\psi \ell'\nu_{\ell'})/dq^2$, $d\mathcal{B}(B_c\to \eta_c \ell'\nu_{\ell'})/dq^2$ and  $d\mathcal{B}(\Lambda_b\to \Lambda_c \ell'\nu_{\ell'})/dq^2$, nevertheless, they have no obvious effects on $d\mathcal{B}(B \to D^* \ell'\nu_{\ell'})/dq^2$  and  $d\mathcal{B}(B \to D \ell'\nu_{\ell'})/dq^2$, since the present accurate experimental measurements of $\mathcal{B}(B \to D^{(\ast)} \ell'\nu_{\ell'})$ give very strongly constraints on the slepton exchange coupling contributions. In addition, $R_{D^*}$  could not give obviously further constraints on  all five  differential branching ratios with $\ell=\ell'$.

\begin{figure}[b]
\begin{center}
\includegraphics[scale=1.1]{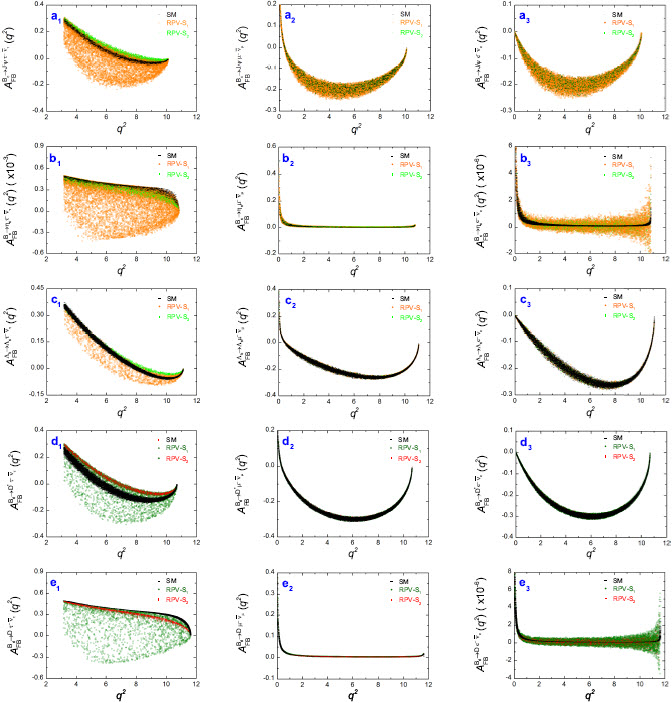}
\end{center}\vspace{-0.8cm}
\caption {The constrained slepton exchange coupling effects on the normalized forward-backward asymmetries of the exclusive semileptonic $b\to c \ell^-\bar{\nu}_\ell$ decays.}
\label{fig:dAFB}
\end{figure}
Fig. \ref{fig:dAFB} displays the constrained slepton exchange coupling effects on the normalized forward-backward asymmetries of the exclusive semileptonic $b\to c \ell^-\bar{\nu}_\ell$ decays.
Since the SM and RPV predictions of the normalized forward-backward asymmetries in cases of $\ell=\mu$ and $\ell=e$ are quite different, we show them all in Fig. \ref{fig:dAFB}.
For $\ell=\tau$ case, as shown in the first column of  Fig. \ref{fig:dAFB}, the significant effects on all five normalized forward-backward asymmetries are allowed in case of $S_1$, the experimental measurement of $R_D^*$ could give very strongly further bounds on these  normalized forward-backward asymmetries. So the measurement of these five normalized forward-backward asymmetries  could test our RPV predictions and further shrink or reveal the parameter spaces of the slepton exchange couplings.
In addition, as shown in Fig. \ref{fig:dAFB} ($b_3$,$e_3$),  the constrained slepton exchange  couplings still provide quite obvious effects
on $A^{B_c\to \eta_c e^-\bar{\nu}_e}_{FB}(q^2)$ and $A^{B_d\to D e^-\bar{\nu}_e}_{FB}(q^2)$, their sign could be changed, nevertheless, both quantities are tiny.

\section{Conclusion}
Motivated by $R_D$, $R_{D^*}$ and $R_{J/\psi}$ anomalies reported by LHCb, BABAR and Belle Collaborations,  we have  studied  RPV supersymmetric effects in $B_c\to J/\psi\ell^- \bar{\nu}_{\ell},\eta_c\ell^- \bar{\nu}_{\ell}$  and
$\Lambda_b \to \Lambda_c \ell^- \bar{\nu}_{\ell}$ decays, which are also induced by $b\to c \ell^- \bar{\nu}_{\ell}$ at quark level. Since  the squark exchange couplings have  tiny effects in these decays, we have only focused on the slepton exchange couplings in this work.

The slepton exchange couplings $\lambda_{i33}\tilde{\lambda}'^*_{i23}$ involve in the exclusive $b\to c \tau^-\bar{\nu}_\tau$ decays. The latest relevant experimental measurements at 95\% CL give obvious bounds on the both moduli and weak phases of $\lambda_{i33}\tilde{\lambda}'^*_{i23}$,  and these  couplings   could explain the recent $R_D$, $R_{D^*}$ and $R_{J/\psi}$ anomalies at the same time.
We have found that, if considering all relevant experimental bounds except for $\mathcal{B}(B_d \to D_d^* \tau^- \bar{\nu}_{\tau})$ and  $\mathcal{R}(D^*)$ at $95\%$ CL,   the constrained slepton couplings have  great effects on  all five (differential) branching ratios with $\ell=\tau$, five ratios of the (differential) branching ratios  and the normalized forward-backward  asymmetries of $\tau^-$.
And the most of branching ratios with $\ell=\tau$ and ratios of the branching ratios  are very sensitive to the both moduli and weak phases of  $\lambda_{i33}\tilde{\lambda}'^*_{i23}$.
The experimental lower limit of $R_{D^*}$ could give further very obvious constraints on slepton exchange couplings.   As for $\mathcal{B}(B_d \to D_d^{\ast } {\tau}^- \bar{\nu}_{\tau})$, which experimental constraints are not used, a narrow range of RPV prediction  could still satisfy  its present measurement.
The upper limit of $R_{J/\psi}$, $R_{\eta_c}$ and $R_{\Lambda_c}$ could be increased by 10\%, 112\% and 24\%, respectively,  from their upper limits of the SM predictions.
The upper limit of RPV prediction of $R_{J/\psi}$ is  1.5$\sigma$ away from its experimental measurement.

The slepton exchange couplings $\lambda_{i11}\tilde{\lambda}'^*_{i23}$ and  $\lambda_{i22}\tilde{\lambda}'^*_{i23}$ involve in the exclusive $b\to c e^-\bar{\nu}_e, s e^+ e^-$ and $b\to c \mu^-\bar{\nu}_\mu,s \mu^+ \mu^-$ decays, respectively. The constrained couplings  of $\lambda_{i11}\tilde{\lambda}'^*_{i23}$ and  $\lambda_{i22}\tilde{\lambda}'^*_{i23}$ from  the exclusive  $b\to s e^+ e^-,s \mu^+ \mu^-$ decays have quite small effects on the branching ratios and their ratios of the exclusive semileptonic $b\to c e^-\bar{\nu}_e$ and $b\to c \mu^-\bar{\nu}_\mu$ decays, nevertheless, the constrained $\lambda_{i11}\tilde{\lambda}'^*_{i23}$ couplings still have obviously effects on the  normalized forward-backward asymmetries of  $B_c\to \eta_ce^- \bar{\nu}_{e}$ and $B_d\to D e^-\bar{\nu}_e$, but both $A_{FB}^{B_c\to \eta_ce^- \bar{\nu}_{e}}(q^2)$ and $A^{B_d\to D e^-\bar{\nu}_e}_{FB}(q^2)$ are tiny.

The large amount of data is expected in the near future from LHCb and BELLE II, and the precise measurements of the ratios of the branching ratios and the  normalized forward-backward asymmetries of the exclusive semileptonic $b\to c \tau \nu_\tau$ decays would enable us to  test our RPV predictions and further shrink or reveal the parameter spaces of the slepton exchange couplings.

\section{Acknowledgments}

This work is supported by the National Natural Science Foundation of China under Contract No.11047145, Nanhu Scholars Program and the
High Performance Computing Lab of Xinyang Normal University.

\begin{appendix}

\section{Appendix}

\subsection{Formulae of the $B_{q} \to M \ell^- \bar{\nu}_\ell$ decays}
\label{A:B}
The hadronic matrix elements for $B_q \to P/V$ transition can be parameterized by the form factors as
\begin{eqnarray}\label{B2PVtransition1}
<P(p')|\bar{q'}\gamma_{\mu}b|B_q(p)> &=& F_+(q^2)\left[(p+p')_{\mu}-\frac{m^2_{B_q}-m^2_P}{q^2}q_{\mu}\right]+F_0(q^2)\frac{m^2_{B_q}-m^2_P}{q^2}q_{\mu},\nonumber\\
<V(p',\epsilon^{\ast})|\bar{q'}\gamma_{\mu}b|B_q(p)>&=& \frac{2iV(q^2)}{m_{B_q}+m_V}\varepsilon_{\mu\nu\rho\sigma} \epsilon^{\ast\nu}p'^{\rho}p^{\sigma},\nonumber\\
<V(p',\epsilon^{\ast})|\bar{q'}\gamma_{\mu}\gamma_5 b|B_q(p)> &=& 2m_V A_0(q^2)\frac{\epsilon^{\ast}\cdot q}{q^2}q_{\mu}+(m_{B_q}+m_V)A_1(q^2)\left[\epsilon^{\ast}_{\mu}-\frac{\epsilon^{\ast}\cdot q}{q^2}q_{\mu}\right]\nonumber\\
&& -A_2(q^2)\frac{\epsilon^{\ast}\cdot q}{(m_{B_q}+m_V)}\left[(p+p')_{\mu}-\frac{m^2_{B_q}-m^2_V}{q^2}q_{\mu}\right],
\end{eqnarray}
where $q=p-p'$ is the momentum transfer, $F_+,F_0$ and $V,A_0,A_1,A_2$ are the form factors of $B_q\to P$ and $B_q\to V$ transitions, respectively.
Noted that, in our numerical results, we take the $B \to D/D^{\ast}$ form factors from Refs. \cite{Caprini:1997mu,Amhis:2016xyh} and the $B_c \to \eta_c,J/\psi$ form factors from Refs. \cite{Wen-Fei:2013uea}.

The double differential branching ratios of  $B_{q} \to P \ell^- \bar{\nu}_\ell$ decays can be represented as
\begin{eqnarray}
\frac{d\mathcal{B}(B_{q} \to P \ell^- \bar{\nu}_\ell)}{dq^2d\cos{\theta_\ell}}&=&\frac{G_F^2|V_{cb}|^2\tau_{B_q}|\vec{p}_P|q^2}{128\pi^3m_{B_q}^2}\left(1-\frac{m_\ell^2}{q^2}\right)^2 \Bigg\{H_0^2 \sin^2{\theta_\ell}\Big(\big|G_V\big|^2+\big|\widetilde G_V\big|^2\Big)\nonumber\\
&&+\frac{m_\ell^2}{q^2}\Big|H_0G_V\cos{\theta_l}-\Big(H_tG_V+\frac{\sqrt{q^2}}{m_\ell}H_SG_S\Big)\Big|^2\nonumber\\
&& +\frac{m_\ell^2}{q^2}\Big|H_0\widetilde G_V\cos{\theta_l}-\Big(H_t\widetilde G_V+\frac{\sqrt{q^2}}{m_\ell}H_S\widetilde G_S\Big)\Big|^2\Bigg\},
\end{eqnarray}
where $|\vec{p}_{M}|\equiv\sqrt{\lambda(m_{B_q}^2,m_{M}^2,q^2)}/2m_{B_q}$ with $\lambda(a,b,c)\equiv a^2+b^2+c^2-2(ab+bc+ca)$.

The double differential branching ratios of  $B_{q} \to V \ell^- \bar{\nu}_\ell$ decays are
\begin{eqnarray}
\frac{d\mathcal{B}(B_{q} \to V \ell^- \bar{\nu}_\ell)}{dq^2d\cos{\theta_l}}&=& \frac{G_F^2|V_{cb}|^2\tau_{B_q}|\vec{p}_V|q^2}{256\pi^3m_{B_q}^2}\left(1-\frac{m_\ell^2}{q^2}\right)^2\Bigg\{2\mathcal{A}_0^2 \Big(\big|G_A\big|^2+\big|\widetilde G_A\big|^2\Big)\sin^2{\theta_\ell}\nonumber\\
&&+(1+\cos^2{\theta_\ell})\Big[\mathcal{A}_{\|}^2\Big(\big|G_A\big|^2+\big|\widetilde G_A\big|^2\Big)+\mathcal{A}_{\perp}^2\Big(\big|G_V\big|^2+\big|\widetilde G_V\big|^2\Big)\Big]\nonumber\\
&&-4\cos{\theta_\ell}Re\Big[\mathcal{A}_{\|}\mathcal{A}_{\perp}\Big(G_AG^*_V-\widetilde G_A\widetilde G^*_V\Big)\Big]\nonumber\\
&&+\frac{m_\ell^2}{q^2}\sin^2{\theta_\ell}\Big[\mathcal{A}_{\|}^2\Big(\big|G_A\big|^2+\big|\widetilde G_A\big|^2\Big)+\mathcal{A}_{\perp}^2\Big(\big|G_V\big|^2+\big|\widetilde G_V\big|^2\Big)\Big]\nonumber\\
&&+\frac{2m_\ell^2}{q^2}\Big|\mathcal{A}_0G_A \cos{\theta_l}-\Big(\mathcal{A}_tG_A+\frac{\sqrt{q^2}}{m_l}\mathcal{A}_PG_P\Big)\Big|^2\nonumber\\
&&+\frac{2m_\ell^2}{q^2}\Big|\mathcal{A}_0\widetilde G_A\cos{\theta_l}-\Big(\mathcal{A}_t\widetilde G_A+\frac{\sqrt{q^2}}{m_l}\mathcal{A}_P \widetilde G_P\Big)\Big|^2\Bigg\}.
\end{eqnarray}

\subsection{Formulae of the $\Lambda_b \to \Lambda_c \ell^- \bar{\nu}_\ell$ decays }
\label{A:Lambdab}
The  hadronic matrix elements for  $\Lambda_b \to \Lambda_c \ell^- \bar{\nu}_\ell$ transition can be parameterized as \cite{Shivashankara:2015cta}
\begin{eqnarray}
<\Lambda_c(p_2,\lambda_2)|\bar{c}\gamma_\mu b|\Lambda_b(p_1,\lambda_1)>&=&\bar u_2(p_2,\lambda_2)[f_1(q^2)\gamma_{\mu}+if_2(q^2){\sigma}_{\mu\nu}q^{\nu}+f_3(q^2)q_{\mu}]u_1(p_1,\lambda_1),
\label{Lb2Lc1}\nonumber\\
<\Lambda_c(p_2,\lambda_2)|\bar{c}\gamma_\mu \gamma^5 b|\Lambda_b(p_1,\lambda_1)>&=&\bar u_2(p_2,\lambda_2)[g_1(q^2)\gamma_{\mu}+ig_2(q^2){\sigma}_{\mu\nu}q^{\nu}+g_3(q^2)q_{\mu}]\gamma_{5}u_1(p_1,\lambda_1),\nonumber\\
<\Lambda_c(p_2,\lambda_2)|\bar{c} b|\Lambda_b(p_1,\lambda_1)>&=&\bar u_2(p_2,\lambda_2)\left[f_1(q^2) \frac{\not q}{m_b-m_c}+f_3(q^2)\frac{q^2}{m_b-m_c}\right]u_1(p_1,\lambda_1),\nonumber\\
<\Lambda_c(p_2,\lambda_2)|\bar{c}\gamma^5 b|\Lambda_b(p_1,\lambda_1)>&=&\bar u_2(p_2,\lambda_2)\left[-g_1(q^2)\frac{\not q}{m_b+m_c}-g_3(q^2)\frac{q^2}{m_b+m_c}\right]\gamma_{5}u_1(p_1,\lambda_1),\nonumber
\end{eqnarray}
where $q=(p_1-p_2)$, $\sigma_{\mu\nu}=i[\gamma_{\mu},\gamma_{\nu}]/2$, $\lambda_i$ is the helicity of baryons, and $f_i(q^2)$, $g_i(q^2)$ are $\Lambda_b \to \Lambda_c$ form factors.
And the form factors are taken from Ref. \cite{Detmold:2015aaa} in our results.

The double differential branching ratios of $\Lambda_b \to \Lambda_c \ell^- \bar{\nu}_\ell$ can be written as \cite{Shivashankara:2015cta}

\begin{equation}\label{Lb2Lc9}
\frac{d\mathcal{B}(\Lambda_b \to \Lambda_c \ell^- \bar{\nu}_\ell)}{dq^2d\cos{\theta_\ell}}=\frac{G_F^2|V_{cb}|^2\tau_{\Lambda_b}|\vec{p}_{\Lambda_c}|q^2}{512\pi^3m^2_{\Lambda_b}}\left(1-\frac{m_\ell^2}{q^2}\right)^2\left[A_1+\frac{m_\ell^2}{q^2}A_2+2A_3+\frac{4m_\ell}{\sqrt{q^2}}A_4\right],
\end{equation}
where $|\vec{p}_{\Lambda_c}|=\sqrt{\lambda(m^2_{\Lambda_b},m^2_{\Lambda_c},q^2)}/{2m_{\Lambda_b}}$, and
\begin{eqnarray}\label{Lb2Lc10}
A_1 &=& 2\sin^2{\theta_\ell}\Big(\big|H_{\frac{1}{2} 0}\big|^2+\big|H_{-\frac{1}{2} 0}\big|^2\Big)+\big(1-\cos{\theta_\ell}\big)^2\big|H_{\frac{1}{2} 1}\big|^2+(1+\cos{\theta_\ell})^2\big|H_{-\frac{1}{2} -1}\big|^2,\nonumber\\
A_2 &=&  2\cos^2{\theta_\ell}\Big(\big|H_{\frac{1}{2} 0}\big|^2+\big|H_{-\frac{1}{2} 0}\big|^2\Big)+\sin^2{\theta_\ell}\Big(\big|H_{\frac{1}{2} 1}\big|^2+\big|H_{-\frac{1}{2}-1}\big|^2 \Big)\nonumber\\
&&+2\Big(\big|H_{\frac{1}{2} t}\big|^2+\big|H_{-\frac{1}{2} t}\big|^2\Big)-4\cos{\theta_\ell}Re\Big[H_{\frac{1}{2} t}H_{\frac{1}{2} 0}^{\ast}+H_{-\frac{1}{2} t}H_{-\frac{1}{2} 0}^{\ast}\Big],\nonumber\\
A_3 &=& \Big|H^{SP}_{\frac{1}{2} 0}\Big|^2+\Big|H^{SP}_{-\frac{1}{2} 0}\Big|^2,\nonumber\\
A_4 &=& -\cos{\theta_\ell}Re\Big[H_{\frac{1}{2} 0}H^{SP *}_{\frac{1}{2} 0}+H_{-\frac{1}{2} 0}H^{SP*}_{-\frac{1}{2} 0}\Big]+Re\Big[H_{\frac{1}{2} t}H^{SP*}_{\frac{1}{2} 0}+H_{-\frac{1}{2} t}H^{SP*}_{-\frac{1}{2} 0}\Big].
\end{eqnarray}

\end{appendix}

\end{document}